# ICARUS and Status of Liquid Argon Technology

**Dorota Stefan on behalf of the ICARUS Collaboration**
Laboratori Nazionali del Gran Sasso dell'INFN,
S.S 17 BIS km. 18.910 67010 Assergi L'Aquila, Italy

E-mail: dorota.stefan@lngs.infn.it

**Abstract**. ICARUS is the largest liquid argon TPC detector ever built (~600 ton LAr mass). It operates underground at the LNGS laboratory in Gran Sasso. It has been smoothly running since summer 2010, collecting data with the CNGS beam and with cosmics. Liquid argon TPCs are really "electronic bubble chamber" providing a completely uniform imaging and calorimetry with unprecedented accuracy on massive volumes. ICARUS is internationally considered as a milestone towards the realization of the next generation of massive detectors (~tens of ktons) for neutrino and rare event physics. Results will be presented on the data collected during 2010 with the detector at LNGS.
*Contribution to NUFACT 11, XIIIth International Workshop on Neutrino Factories, Super beams and Beta beams, 1-6 August 2011, CERN and University of Geneva (Submitted to IOP conference series).*

## 1. Introduction

Currently the Liquid Argon Time Projection Chamber (LAr-TPC) technology [1] is of great interest due to its potential in investigation of the neutrino properties and rare events such as proton decay or dark matter. The sensitivity is around 4 times more with respect to the similar measurement performed in the water Cherenkov detector of the same mass. The ICARUS detector is the largest liquid Argon (LAr) detector which has ever been built, with mass of 600 ton (T600). The experience gained from the detector construction and operation and also data analysis should be further used in the future LAr detectors. The advantage of the LAr TPC is its excellent spatial and calorimetric resolution which makes possible a perfect visualization of the charged particles tracks. Such a detection technique is often compared to the bubble chamber filled with heavy liquids, but with the electronic readout. In order to take advantage of the modern experimental techniques, the track reconstruction and data analysis should be fully automatized. This requires studying new reconstruction methods and solving many algorithmic problems, which are similar to those from automatic image processing and pattern recognition.

## 2. The ICARUS experiment

The ICARUS detector [2] is now installed in the Hall B of the Gran Sasso underground National Laboratory (LNGS) of Istituto Nazionale di Fisica Nucleare (INFN), shielded against cosmic rays by about 1400 meters of rock. The detector is running smoothly under stable conditions starting from October 1$^{st}$ 2010. It is simultaneously collecting self-triggered events of cosmic rays and neutrino interactions associated with the CNGS neutrino beam, focusing on neutrino oscillation search and in particular on $\nu_\mu \rightarrow \nu_\tau$ appearance. The trigger system and liquid Argon purity measurement of the detector are very important issues, described in [2].

The detection principle is following. Charged particles, generated for example by a neutrino interaction in LAr, produce ionization along their path. Thanks to the low transverse diffusion of charge in LAr and to its exceptionally high purity, electrons are drifted along the lines of the electric field applied to the LAr volume and, are projected onto the anode (TPC).

Anode is made of three parallel planes of wires, 3 mm apart and 3 mm of wire pitch, facing the drift path of 1.5 m. By appropriate voltage biasing, the first two planes (Induction-1, Induction-2) provide signals in a non-destructive way, whereas the ionization charge is finally collected by the last one (Collection plane).

Wires are oriented on each plane at a different angle ($0^o$, $+60^o$, $-60^o$) with respect to the horizontal direction. Therefore, combining the wire coordinates on each plane at a given drift time, a three-dimensional image of the ionizing event can be reconstructed.

The measurement of the absolute time of the ionizing event is combined with the electron drift velocity information, and provides the position of the track along the drift coordinate. The determination of the absolute time of the ionizing event is accomplished by the prompt detection of the scintillation light produced in LAr by charged particles.

## 3. Approach to the event reconstruction

In the event reconstruction two types of objects are distinguished: tracks and cascades. The scheme of the reconstruction procedure is presented on the figure 1. First, the signals (hits) on all three independent wire planes have to be found. Next, only in the Collection wire planes, the fitting procedure to the signal shape is applied to obtain the charge deposition on wires. At this stage, the rough estimation of calorimetric measurement can be provided.

### 3.1. Spatial reconstruction

In order to obtain precise calorimetric information for tracks, the 3D reconstruction is necessary. For this purpose, first, the 2D hit clusters are found separately for two wire planes. Then Polygonal Line Algorithm (PLA) [3] is used to form the 2D trajectories and sort hits along them. Finally, 3D track reconstruction is performed by matching hits from two corresponding 2D trajectories, while third wire plane is used in ambiguous topologies. Matching is based on both the same electron drift time and order of hits sorted with PLA.

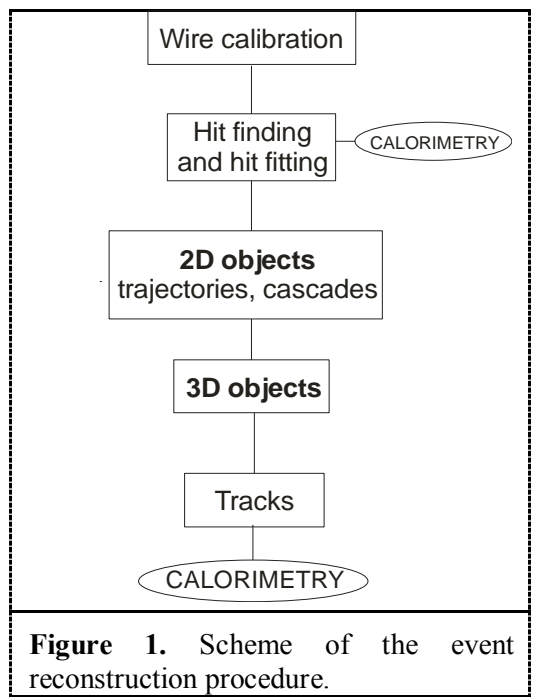

**Figure 1.** Scheme of the event reconstruction procedure.

## 3.2. Calorimetric reconstruction

The ionization charge is measured at the collection wire plane. The energy $E$ associated to a given hit is:

$$E = \frac{CW}{R} Q e^{\frac{t-t_0}{\tau}}$$

where: $C$ – the calibration factor; $W$ – the average energy needed for the creation of an electron-ion pair; $R$ – the electron-ion recombination factor; $Q$ – the primary deposited ionization charge; $t$ – the electron drift time; $t_0$ – time provided by the photomultiplier trigger; $\tau$ - the electron lifetime in LAr.

The recombination factor $R$ depends on the absorber medium, on the applied electric field and on the density of the released charge. In case of the electromagnetic cascades, the constant mean recombination factor is applied for all hits while for the tracks precise dependence of recombination on $dE/dx$ can be modeled by Birk's law, which has been applied to LAr detectors in [4]:

$$\frac{dE}{dx} = \frac{dQ}{A - \frac{k}{\varepsilon\rho} \times \frac{dQ}{dx}}$$

where: $dE/dx$ – a real particle ionization loss per track length unit; $A$ – the normalization factor (~0.8); $k$ – the Birks coefficient accounting for recombination (~0.05 (kV/cm)(g/cm$^2$)); $\varepsilon$ – electric field (~0.5 kV/cm); $\rho$ - the liquid argon density (~1.4 g/cm$^3$).

For a long muon track escaping the detector, the momentum is determined exploiting the multiple scattering along the track, studying the displacements with respect to a straight line. The procedure, implemented as a Kalman filter technique and validated on cosmic ray stopping muons, allows a resolution Δp/p that can be as good as 10%, depending mainly on the track length [5].

## 3.3. Particle identification

Successful particle identification depends largely on the efficiency of the spatial and calorimetric reconstruction. In case of particles stopping in the detector the energy deposition per track length unit as a function of the deposited energy of the residual range allows to identify particles species (muon, pion, kaon, proton). The information from 3D track reconstruction is the input to a neural network based algorithm for particle identification [6]. A high efficiency has been achieved, above 85%, for particles with deposited energy around 30 MeV and higher (which corresponds to the length of at least 6 cm).

Electrons are fully identified by the characteristic electromagnetic showering; they are well separated from $\pi^0$ via $dE/dx$ signal around the conversion point, conversion distance and $\pi^0$ invariant mass measurement at the level of 10$^{-3}$. This feature guarantees a powerful identification of the CC electron neutrino interactions, while rejecting NC interactions to a negligible level.

## 4. Data analysis

The ICARUS analysis framework allows a complete kinematical reconstruction of the neutrino interactions. The analysis of events collected during 2010 is on-going. They are very important in order to develop and test tools for the automatized reconstruction procedure. Analysis of the muon momentum in the CC interaction events, the study of the muon directions, and also measurement of the $\nu_\mu$ energy show good agreement with Monte Carlo simulation performed in FLUKA [7].

One simple example of the fully reconstructed NC neutrino event with production of η particle is presented in figure 2. There are two electromagnetic cascades pointing to the primary vertex and tracks which were reconstructed and analysed. The conversion distances of cascades are measured to be 26 cm and 12 cm respectively. The associated invariant mass m$_{\gamma\gamma}$* = 512 ± 48 MeV is compatible with the η mass. The ionization signal in 2.5 cm of the initial part of the cascade amounts to be 4.3 and 5.1 m.i.p. (minimum ionizing particle) on average. This is clear signature of a pair conversion confirming the expected e/γ identification capabilities of the detector. The trajectory visible in both projections, Collection and Induction-2, is the result of the chain of particle decays. Starting from primary vertex they are following: pion, muon and electron/positron. Those tracks were reconstructed in 3D and precise energy measurement was provided. The total momentum of the event is around 2.2 GeV.

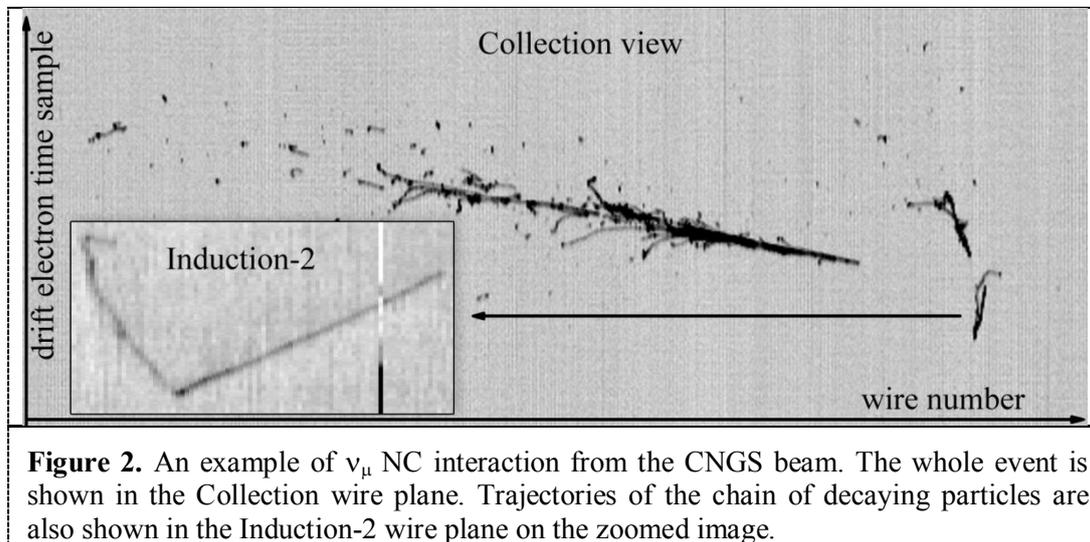

**Figure 2.** An example of $\nu_\mu$ NC interaction from the CNGS beam. The whole event is shown in the Collection wire plane. Trajectories of the chain of decaying particles are also shown in the Induction-2 wire plane on the zoomed image.

## 5. Summary
The ICARUS detector is so far the biggest LAr detector ever built. It has been successfully installed in the Gran Sasso underground laboratory and it is presently collecting data. Thanks to very good spatial-calorimetric resolution and its underground operation a LAr detector is a perfect tool in research of neutrino physics and proton decay. Therefore the experience gained from the ICARUS experiment will be very valuable for the future research in these fields. As far as reconstruction and data analysis are concerned, one should stress that collected neutrino interaction events are used in order to develop and test automatic event reconstruction, allowing batch analysis of the data collected in long run periods.